\documentclass{optica-article}

\journal{opticajournal} % for journals or Optica Open

\articletype{Research Article}

\usepackage{lineno, soul, color}
\begin{document}

\title{HAMscope: a snapshot Hyperspectral Autofluorescence Miniscope for real-time molecular imaging}

\author{Alexander Ingold,\authormark{1} Richard G. Baird,\authormark{1} Dasmeet Kaur,\authormark{2} Nidhi Dwivedi,\authormark{2} Reed Sorenson,\authormark{3} Leslie Sieburth,\authormark{3}  Chang-Jun Liu,\authormark{2} and Rajesh Menon\authormark{1,*}}

\address{\authormark{1}Dept. of Electrical \& Computer Engineering, University of Utah, Salt Lake City UT 84112, USA\\
\authormark{2}Biological, Environmental and Climate Sciences Department, Brookhaven National Laboratory, 98 Rochester St, Upton, NY 11973, USA\\
\authormark{3}Dept. of Biology, University of Utah, Salt Lake City UT 84112, USA
}

\email{\authormark{*}rmenon@eng.utah.edu} %% email address is required; see note below about the corresponding author designation

% use {asbstract*} to suppress the copyright line. Copyright information will be added in production

\begin{abstract*} 
We introduce HAMscope, a compact, snapshot hyperspectral autofluorescence miniscope that enables real-time, label-free molecular imaging in a wide range of biological systems. By integrating a thin polymer diffuser into a widefield miniscope, HAMscope spectrally encodes each frame and employs a probabilistic deep learning framework to reconstruct 30-channel hyperspectral stacks (452–703 nm) or directly infer molecular composition maps from single images. A scalable multi-pass U-Net architecture with transformer-based attention and per-pixel uncertainty estimation enables high spatio-spectral fidelity (mean absolute error $\sim$0.0048) at video rates. While initially demonstrated in plant systems, including lignin, chlorophyll, and suberin imaging in intact poplar and cork tissues, the platform is readily adaptable to other applications such as neural activity mapping, metabolic profiling, and histopathology. We show that the system generalizes to out-of-distribution tissue types and supports direct molecular mapping without the need for spectral unmixing. HAMscope establishes a general framework for compact, uncertainty-aware spectral imaging that combines minimal optics with advanced deep learning, offering broad utility for real-time biochemical imaging across neuroscience, environmental monitoring, and biomedicine. 
\end{abstract*}

%%%%%%%%%%%%%%%%%%%%%%%%%%  body  %%%%%%%%%%%%%%%%%%%%%%%%%%
\section{Introduction}
Fluorescence microscopy has become a cornerstone of biological imaging, yet its application to plant biology—especially in field settings—remains limited by the widespread dependence on exogenous fluorescent labels. These labels often require complex delivery mechanisms, are susceptible to photobleaching, and may perturb the native biochemical or physiological state of the tissue \cite{hickey2021fluorescence}. Moreover, many fluorescent dyes are not optimized for use in intact plants, where cell walls, cuticles, and autofluorescent biomolecules introduce substantial background and complicate interpretation \cite{ckurshumova2011glow}. In ecological or agricultural contexts, delivering labels across large plant populations or into vascular tissues in situ is impractical and disruptive, particularly under field conditions where sterile access, reagent stability, and environmental control are lacking.

Label-free imaging based on endogenous autofluorescence provides a promising alternative, particularly in plants where biomolecules such as lignin, chlorophyll, flavins, and phenolics exhibit rich intrinsic fluorescence across the UV-visible spectrum \cite{Donaldson2020}. These autofluorescence signatures report on biochemical composition, stress responses, and developmental states without invasive labeling or external reagents. Hyperspectral imaging further enables the decomposition of these signals into spectrally resolved components, allowing biomolecular classification and mapping at the tissue or cellular level \cite{saric2022applications}. However, current hyperspectral autofluorescence platforms are typically benchtop systems that are bulky, expensive, and sensitive to environmental perturbations. These limitations have prevented the widespread use of hyperspectral autofluorescence imaging in natural ecosystems and agricultural settings, where capturing dynamic processes—such as embolism formation, senescence, and pathogen response—is critically important\cite{domozych2012quest}. At present, monitoring these phenomena typically relies on labor-intensive manual sampling or low-resolution aerial imaging,\cite{maes2019perspectives} both of which lack the spatial and biochemical resolution required for detailed analysis.

Here, we introduce a compact, field-deployable solution for label-free biochemical imaging by transforming the widely adopted miniaturized widefield microscope (the \emph{miniscope}) \cite{aharoni2019all} into a snapshot Hyperspectral Autofluorescence Miniscope, {\it i.e., the HAMscope}. This is achieved by integrating a thin polymer diffuser at the image plane and implementing a machine-learning pipeline that reconstructs 30 spectral channels from each raw sensor frame. A standard UV-LED source provides excitation. With these minimal hardware modifications, our system preserves the native spatial resolution and full frame rate of the underlying CMOS detector while enabling dense spectral capture in a single shot. Crucially, the modified miniscope retains its original size and weight, supporting low-cost, field-ready deployment. Full system specifications and implementation details are provided in the Supplementary Information and on the project Github page \cite{Hamscope_github}.   

We note that snapshot hyperspectral imaging systems typically fall into two main categories. The first extends conventional color filter arrays (e.g., RGB) to a greater number of spectral channels.\cite{vunckx2021accurate} While conceptually simple, this approach suffers from limited scalability—rarely exceeding six channels—and imposes significant trade-offs in light throughput and spatial resolution. The second class employs dispersive optics, either planar or volumetric, to encode spectral information across spatially overlapping measurements, which are then computationally demixed. Dispersive elements placed in Fourier configurations can improve spectral resolution, but often require larger optical paths\cite{bacca2023computational} and may rely on coded apertures, substantially reducing light efficiency.\cite{arce2013compressive} More compact variants position thin dispersive layers directly at the image plane, enabling simpler, low-loss architectures.\cite{wang2017computational,wang2015ultra,majumder2024hd,Bian2024} Across all approaches, spectral reconstruction reduces to solving a high-dimensional, ill-conditioned inverse problem, typically addressed through regularization\cite{wang2014computational,Monakhova2020} or data-driven machine learning.\cite{huang2022spectral} Building on this rich body of work, we extend computational hyperspectral imaging to the ultra-compact miniscope platform and demonstrate snapshot autofluorescence imaging of living plant tissues.

\section{The HAMscope}
We modified the open-source Miniscope V4 platform (Open Ephys) to enable snapshot hyperspectral autofluorescence imaging. This miniscope uses an Aptina MT9V032 CMOS sensor with 608 $\times$ 608 pixels (6 $\mu$m pixel pitch) and spectral sensitivity spanning 452–703 nm \cite{miniscope-cmos-pcb}. To excite autofluorescence, we integrated a 385 nm ultraviolet LED (Lumileds, LHUV-0385-A045), selected to match the absorption features of key plant biomolecules. The excitation light was bandpass filtered (Chroma, ET385/70x, $385\pm35$nm) and paired with a long-pass emission filter (Chroma, ET425lp) and a dichroic mirror (Chroma, T412lpxt, 412 nm cut-on), allowing for efficient capture of both blue–green autofluorescence (e.g., lignin; excitation peak at 380 nm, emission peak $\sim$ 475 nm) \cite{lignin_autofluorescence} and far-red signals from chlorophyll (650–750 nm) \cite{Lamb2018}.

\begin{figure}[htb!]
\centering
\includegraphics[width=\textwidth]{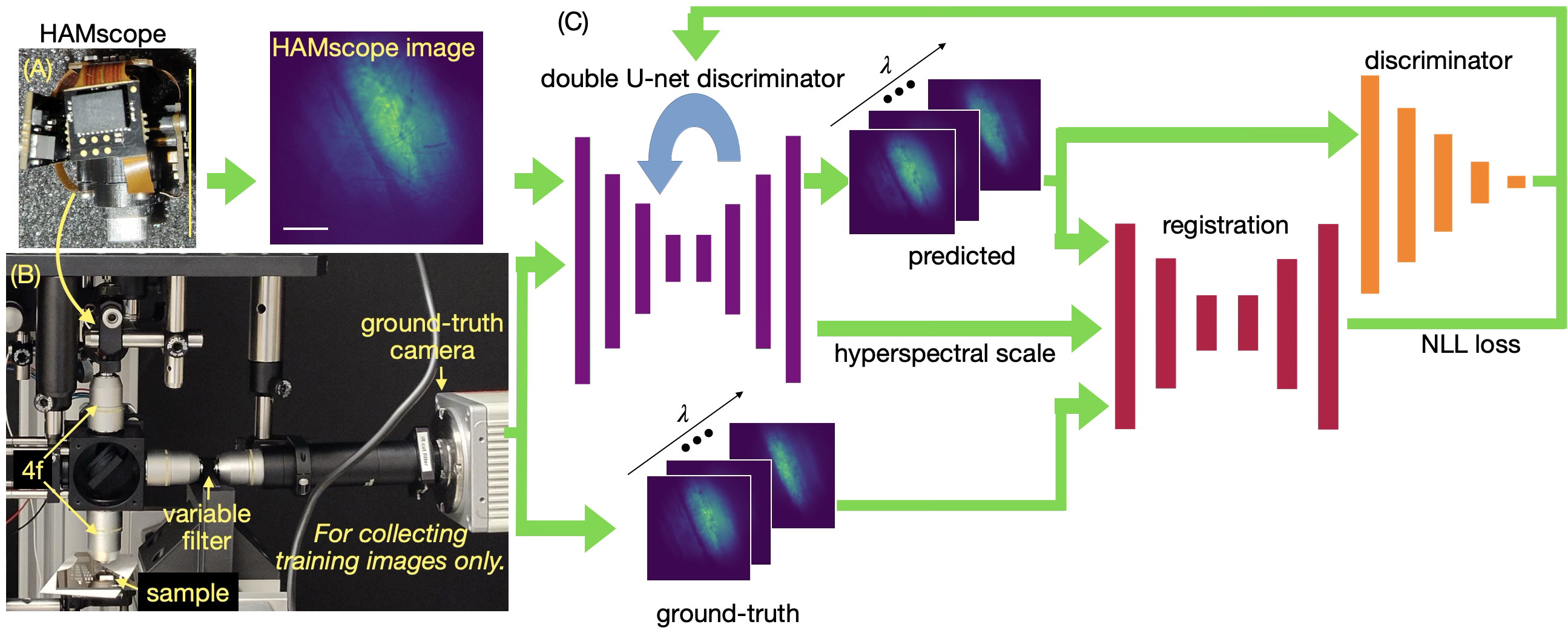}
\caption{The HAMscope and reference imaging setup used for collecting training data. (A) The Miniscope V4 is modified by incorporating a thin polymer diffuser (not visible) at the image plane to enable spectral encoding (scale bar = 22 mm). A UV-LED excitation source, coupled with custom excitation and emission filters, supports label-free autofluorescence imaging of plant biomolecules such as lignin and chlorophyll. (B) A reference benchtop microscope configured in a 4F layout acquires ground-truth hyperspectral images using a tunable narrow-bandpass filter mounted on a motorized scanning stage. Reference and miniscope data are recorded simultaneously as a paired dataset. (C) The model takes one HAMscope image (scale bar = $200\thinspace\mu$m) as input. A multi-U-Net generator produces a predicted hyperspectral stack or a compound-specific classification map derived from the spectral data. The generator also produces a per-pixel uncertainty map, enabling confidence estimation without ground-truth. Following the terminology in \cite{weigert2018}, we refer to this as the hyperspectral scale. Paired input–output datasets (HAMscope and ground-truth hyperspectral images) are used to train the model. A Laplacian negative log-likelihood (NLL) loss guides the probabilistic predictions, while an adversarial discriminator further encourages perceptual realism in the output.}\label{fig:miniscope}
\end{figure}

To encode spectral information in each frame, we inserted a thin polymer diffuser between the emission filter and the CMOS sensor. This element introduces spatially localized, spectrally dependent modulation, while preserving image sharpness. The diffuser angle was empirically measured at 0.046$^{\circ}$, providing sufficient spectral encoding with minimal degradation of spatial resolution. Full details of the diffuser configuration and its characterization are provided in Supplementary Figs. S1 and S2.

The miniscope used lens configuration 1, yielding a field-of-view of $\sim$940 $\mu$m, as specified in the Miniscope V4 documentation \cite{AharoniLabMiniscopeV4}. To enable rapid volumetric imaging, we integrated an electrowetting variable-focus lens (Varioptic A25H), which allowed for continuous axial scanning over a depth range of up to 400 $\mu$m without mechanical movement. The variable working distance is essential for field deployment, as it eliminates the need for precision stages and allows the system to be mounted directly onto a branch using a simple 3D-printed clamp adapter\cite{Hamscope_github}. Power and data were transmitted through a single coaxial cable, interfacing with the DAQ PCB module (Labmaker) for synchronized control and acquisition. The miniscope housing was fabricated via 3D printing from STL files provided in the open-source miniscope V4 repository, using a 0.2 mm nozzle (Bambu Labs, A1M) for high-fidelity component production. Figure \ref{fig:miniscope}A shows a photograph of the snapshot hyperspectral autofluorescence miniscope, which retains the same compact form factor as a conventional miniscope (also see Fig. S3).

To train the machine learning model for hyperspectral reconstruction, we developed a custom ground-truth imaging system that records paired hyperspectral data for each frame acquired by the HAMscope (Fig. S3). A dual-channel 4f relay ensures that the same optical scene is projected simultaneously to the modified miniscope and the reference hyperspectral microscope (Fig. \ref{fig:miniscope}B). The 4f relay comprises two matched 10× objectives (Amscope, PA10X-V300) mounted on opposite ends of a cage system (Thorlabs, C4W). A shortpass dichroic mirror (400 nm cutoff; Newport, 10SWF-400-B) positioned at the Fourier plane directs light toward the reference path. To maximize optical throughput during HAMscope image capture, the dichroic is mounted on a motorized rotation stage and can be retracted from the beam path. The mount is custom 3D-printed and driven by a NEMA 17 stepper motor (Stepperonline), a TMC2208 driver (Dorhea), and a Raspberry Pi 5 (Adafruit) powered by a precision supply (BK Precision, Model 1601). 

Spectral selection in the reference microscope is performed using a linear variable bandpass filter (LVBF; Edmund Optics, 88-365) positioned between two additional 10$\times$ objectives. By translating the LVBF across a focused beam spot, we obtain narrow spectral slices ($\sim$10 nm bandwidth) across 30 channels spanning 452–703 nm. The filter position is controlled with a motorized linear stage (RATTMMOTOR, ZBX80, 100 mm stroke), and spectral calibration is performed with a fiber-coupled spectrometer (Ocean Optics, JAZ A1100, fig. S4). The image was captured using a sCMOS sensor (Hamamatsu, ORCA-Flash4.0 LT) using a 160 mm focal length tube lens. Samples are mounted below the 4f relay on a three-axis motorized stage (Thorlabs, MLJ050), and trigger signals for filter scanning and dichroic actuation are coordinated using a LabJack U3 controller. The system runs on a Windows 10 PC (Intel i7-3770, 32 GB RAM). A complete parts list is provided in Table S3.

The miniscope images were downsampled from 608 $\times$ 608 to 512 $\times$ 512 pixels to standardize input dimensions for training. Hyperspectral ground-truth images were automatically cropped, spatially registered to the corresponding HAMscope frame, and interpolated to the same resolution. Prior to training, all images were normalized to the [0,1] range by dividing by their respective bit depths—256 for 8-bit miniscope images and 65,536 for 16-bit hyperspectral stacks. This fixed scaling preserves relative intensity across channels and frames, avoiding distortions introduced by per-image min/max or mean/variance normalization. To further preserve relative channel intensity, a custom normalization function replaced the instance normalization at all layers of the network.

\subsection{Machine learning}
To reconstruct hyperspectral images from raw HAMscope frames, we developed a probabilistic generative adversarial network (GAN) based on a modified multi-U-Net architecture, adapted from the pix2pix framework \cite{isola2018} (see Fig. \ref{fig:miniscope}C). The network was trained using paired datasets: each training sample consisted of a single monochrome image acquired with the HAMscope and a corresponding 30-channel hyperspectral stack obtained from the ground-truth reference microscope. The dataset spans four incisions on four separate branches, with 6248 training images and 100 randomly selected test images. The hyperspectral images were further processed for compound-specific classification into three-channel maps representing lignin, chlorophyll, and other autofluorescent species. The GAN framework comprises two convolutional neural networks: a generator that predicts the hyperspectral image from the monochrome input and a discriminator that learns to distinguish predicted outputs from real measurements. A composite loss function guides training: the discriminator’s adversarial loss encourages realistic image generation, while an L1 loss enforces pixel-level fidelity. To prevent overfitting or artifact hallucination, the discriminator loss was scaled by a factor of 0.005, ensuring it remained subordinate to the L1 term. The discriminator loss term is set to 0, deactivating the discriminator, in all experiments unless otherwise specified. All models were trained for 30 epochs on a single consumer-grade GPU (NVIDIA RTX 3060). A representative loss curve is shown in Fig. S5.

Our probabilistic reconstruction model outputs the predicted mean and standard deviation for each spectral channel. Training was guided by a Laplacian negative log-likelihood (NLL) loss, which promotes accurate estimation of both the mean signal and associated uncertainty \cite{weigert2018}. These uncertainty estimates are especially critical when imaging directly with the HAMscope in field settings, where no ground-truth is available. Decoupling the HAMscope from the 4f relay and tunable filter—used only for training—was essential to realizing a compact, field-deployable system. Unless otherwise noted (e.g., Fig. S6), all models presented in this work are probabilistic.

To quantify spatially and spectrally resolved aleatoric uncertainty, following the terminology in ref. \cite{weigert2018}, we introduced hyperspectral scale maps generated by the probabilistic model for each spectral channel. Scale maps are critical for evaluating the quality of generated images in the field when no ground-truth images are available. To quantify epistemic uncertainty, we calculated a disagreement score between an ensemble of five independently trained probabilistic networks (Fig. S7). For each pixel, we computed a disagreement score defined as the average pairwise Kullback–Leibler (KL) divergence among predicted distributions, normalized by the log of the number of models ({\it i.e.}, mean(KL)/log(5)) \cite{weigert2018}. This score ranges from 0 (perfect agreement) to 1 (maximum disagreement). As a baseline, random noise produced a disagreement score of 1 when evaluated on five sets of 100 image pairs.

To balance model expressiveness with computational efficiency, we developed a scalable multi-pass U-Net architecture that emulates the depth of larger models without increasing the parameter count. In this design, feature maps from the decoder at each resolution level are propagated to the encoder of the subsequent pass via shared skip connections, effectively deepening the network through iterative reuse of representations. The number of passes is set by a user-defined parameter, with marginal performance gains beyond three iterations. This architecture achieves accuracy comparable to a standard U-Net configured with 2 $\times$ more filters per layer (64–128 channels) and 4 $\times$ the total parameters, yet does not significantly increase VRAM usage and remains trainable on a standard NVIDIA RTX 3060 GPU. In contrast, the control model with similar performance required an RTX 3090.

To enhance the representational capacity of the U-Net generator, we incorporated a transformer block into its deeper layers, following the approach of Tripathy {\it et al.} \cite{tripathy2020}. To keep VRAM requirements tractable, the transformer was applied only after downsampling to a 64 $\times$ 64 spatial resolution. A 2D learned positional embedding was used to encode spatial relationships, enabling the model to capture global context while preserving the image’s two-dimensional structure. This positional encoding was added element-wise to the input feature map and passed to the multi-head self-attention mechanism. The attention module received the same encoded input for the query, key, and value matrices, enabling it to learn context-dependent weighting of spatial features. The block employed eight attention heads, followed by a residual connection and layer normalization to stabilize training and emphasize informative features. The output of the transformer was then passed to the decoder path of the U-Net.

In addition to the traditional quadratic transformer, we explored a transformer attention module with linear scaling, know as the spectral attention model \cite{Bian2024}. The linear scaling reduces Vram requirements, allowing the module to apply to all layers of the U-Net. We chose to use the quadratic transformer as the primary option, as it outperforms the linear transformer for the probabilistic hyperspectral and compound identification models, see Table S1.

During training, we used a separate registration U-Net to correct for residual misalignments between the predicted and ground-truth image stacks \cite{kong2021}. This network estimates a deformable vector field that shifts the predicted image and associated error map prior to NLL loss calculation. Importantly, the registration network is used only during training and is not required at inference time.

\section{Results}
\begin{figure}[htb]
\centering
\includegraphics[width=\linewidth]{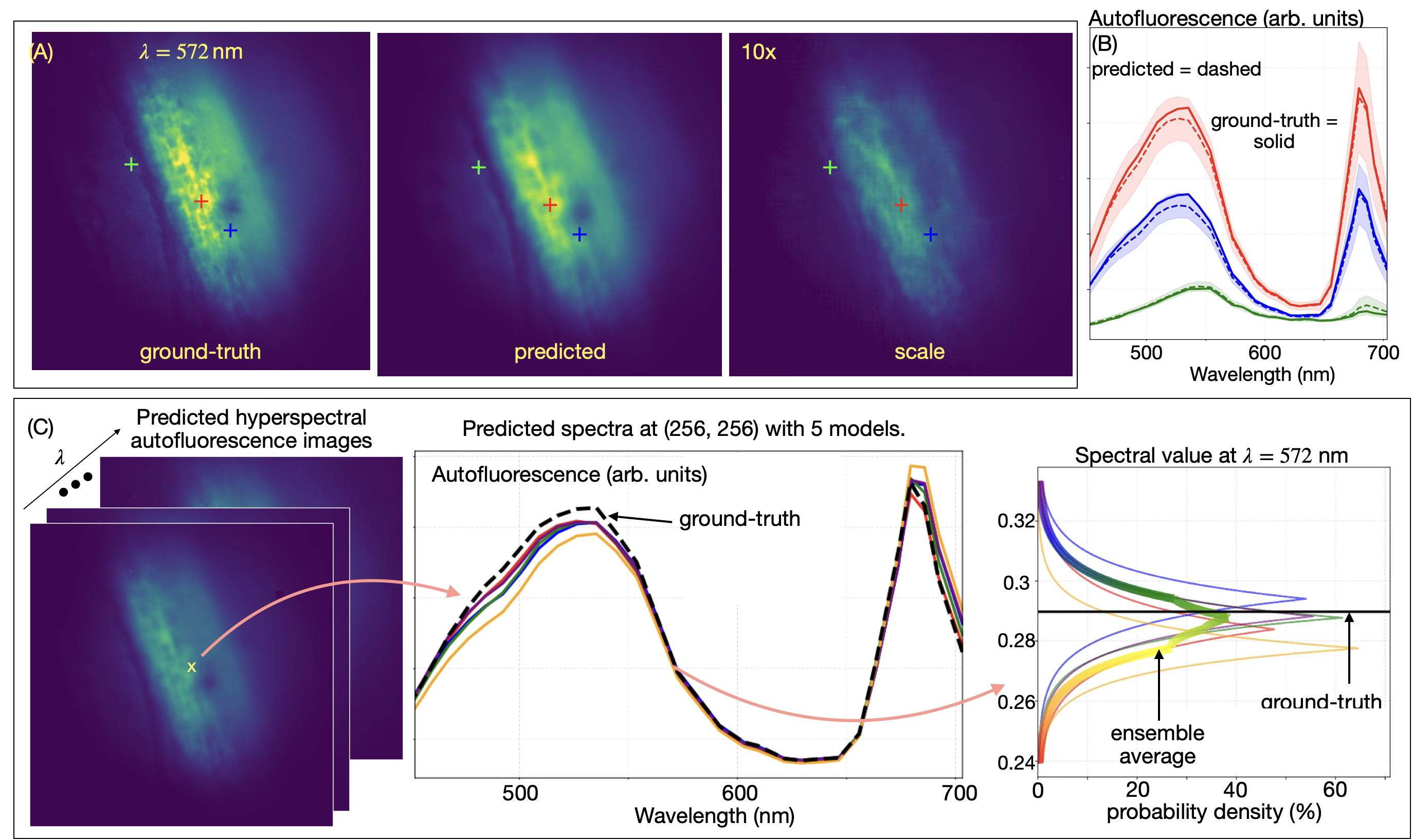}
\caption{Representative results from the HAMscope. (A) Reconstruction of the 572 nm spectral channel for a poplar stem cross-section: ground-truth (left), model prediction (center), and per-pixel uncertainty (right), expressed as the standard deviation from the probabilistic model ensemble. (B) Predicted and ground-truth spectral profiles at three locations (marked by colored “+” symbols), showing close agreement. Solid lines represent ground-truth, dashed lines indicate predictions; shaded regions denote $\pm2$ standard deviations. (C) Left: Full 30-channel hyperspectral image predicted from a single miniscope frame. Center: Predicted spectra at pixel (256, 256) from five independently trained probabilistic models compared with ground-truth. Right: Probability density functions (Laplacian) for the 572 nm channel, with the ground-truth value indicated. The ensemble prediction peaks closely align with the ground-truth, reflecting high model confidence. The results here used a single U-net model with instance normalization. }\label{HSI_core_results}
\end{figure}

A central innovation of our approach is the reconstruction of high-dimensional hyperspectral data from a single monochrome image acquired through a polymer diffuser. The HAMscope reliably predicts all 30 spectral channels from each input frame using the trained probabilistic U-Net. A representative reconstruction of the 572 nm channel is shown in Fig. \ref{HSI_core_results}A, closely matching the corresponding ground-truth. The model accurately recovers complete spectral profiles across all channels as indicated by the spectra from three exemplary points in the image shown in Fig. \ref{HSI_core_results}B. The ground-truth and predicted spectra are shown by solid and dashed lines, respectively. The shading indicates a range of $\pm 2$ standard deviations. The spectral-reproduction error averaged over 100 test images was 6.67\% for the probabilistic single U-net. 

As noted earlier, to quantify both spatially and spectrally resolved aleatoric uncertainty, we adopted the terminology of Weigert et al. \cite{weigert2018} and introduced hyperspectral scale maps, generated by the probabilistic model for each spectral channel. These scale maps provide a principled estimate of local uncertainty and are particularly valuable for evaluating image quality in field settings where ground-truth is unavailable. Representative spectra from three distinct spatial locations (indicated by colored “+” markers in Fig. \ref{HSI_core_results}A) are shown in Fig. \ref{HSI_core_results}B, with shaded bands denoting the $\pm$2 standard deviation range across an ensemble of models, reflecting the model’s predictive confidence. As noted earlier, to further characterize epistemic uncertainty, we computed a disagreement score across five independently trained probabilistic networks (Fig. S2).

To further assess the performance of the hyperspectral miniscope, Fig. \ref{HSI_core_results}C presents a representative 30-channel autofluorescence image acquired from a tangential incision on a living poplar stem from the test dataset. To quantify spectral prediction accuracy, we analyzed the output at a single pixel (coordinates: 256, 256) across five independently trained probabilistic models. The predicted spectra showed strong agreement with the ground-truth across all wavelengths. To evaluate model confidence, we plotted the probability density functions for the 572 nm channel, demonstrating that the ensemble mean closely aligned with the ground-truth value. This consistency across the ensemble confirms the accuracy and robustness of the model’s spectral predictions.

\subsection{Biomolecule mapping}
\begin{figure}[htb!]
\centering
\includegraphics[width=0.7\linewidth]{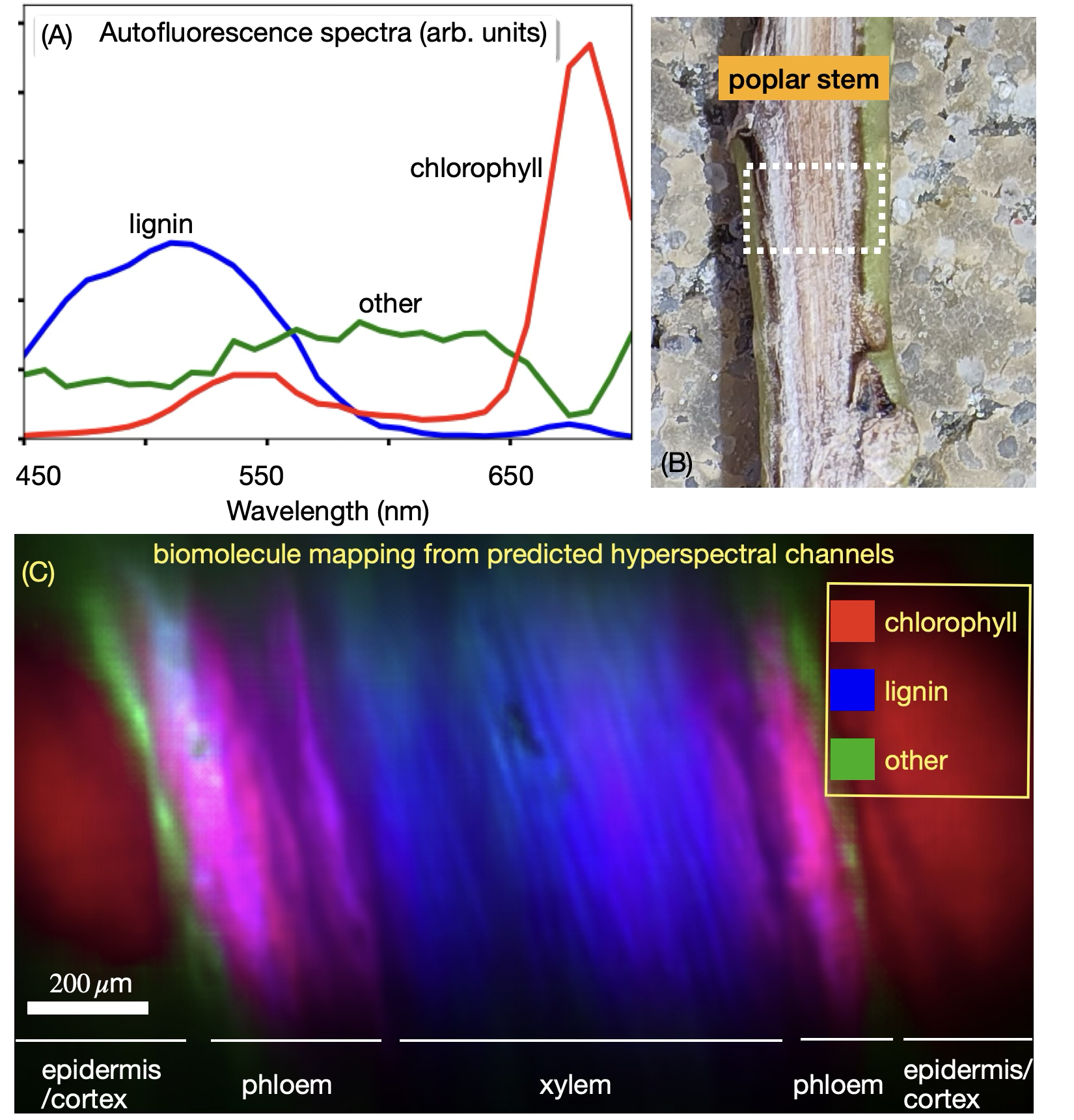}
\caption{Biomolecule mapping from predicted hyperspectral data across a tangential branch incision. (A) Autofluorescence spectra corresponding to lignin, chlorophyll, and other components, extracted from ground-truth hyperspectral images using the PoissonNMF plugin in ImageJ. These spectra were used uniformly across the training set to generate biomolecule maps from the predicted hyperspectral images. (B) Photograph of a tangential incision on a poplar branch, enabling analysis of biomolecular composition as a function of tissue depth. The dashed white box denotes the imaged region. (C) Composite image generated by stitching multiple fields of view, each processed using the PoissonNMF plugin in ImageJ applied to the predicted hyperspectral channels. The resulting color-coded map highlights the spatial distribution of lignin and chlorophyll across anatomical layers. Color legend indicates the spectral identity of each component.} 
\label{spectral_unmixing}
\end{figure}

Hyperspectral imaging captures endogenous molecular contrast by resolving autofluorescent signatures of key biomolecules. To recover this biochemical information from predicted hyperspectral stacks, we performed spectral unmixing using the PoissonNMF plugin in ImageJ \cite{Neher2009}. The data were decomposed into three spectral components, constrained by experimentally validated reference spectra for lignin and chlorophyll \cite{Donaldson2018}. Pixels matching these spectral signatures were assigned to their respective channels, while the remaining signals were aggregated into a third residual “other” component. Unmixing was conducted on tangential branch sections to reveal a view of tissue layers including phloem and xylem (Figs. \ref{spectral_unmixing}A–B). The resulting component maps revealed anatomically consistent spatial patterns, with lignin localized to interior xylem regions and chlorophyll concentrated in outer tissues (Fig. \ref{spectral_unmixing}C). Ground-truth biomolecule mapping is in Fig. S8. Ground-truth region-of-interest analysis further confirmed the correspondence between unmixed spectral features and known tissue structures, including xylem, phloem, and epidermis layers (Fig. S9).

\begin{figure}[htb!]
\centering
\includegraphics[width=\linewidth]{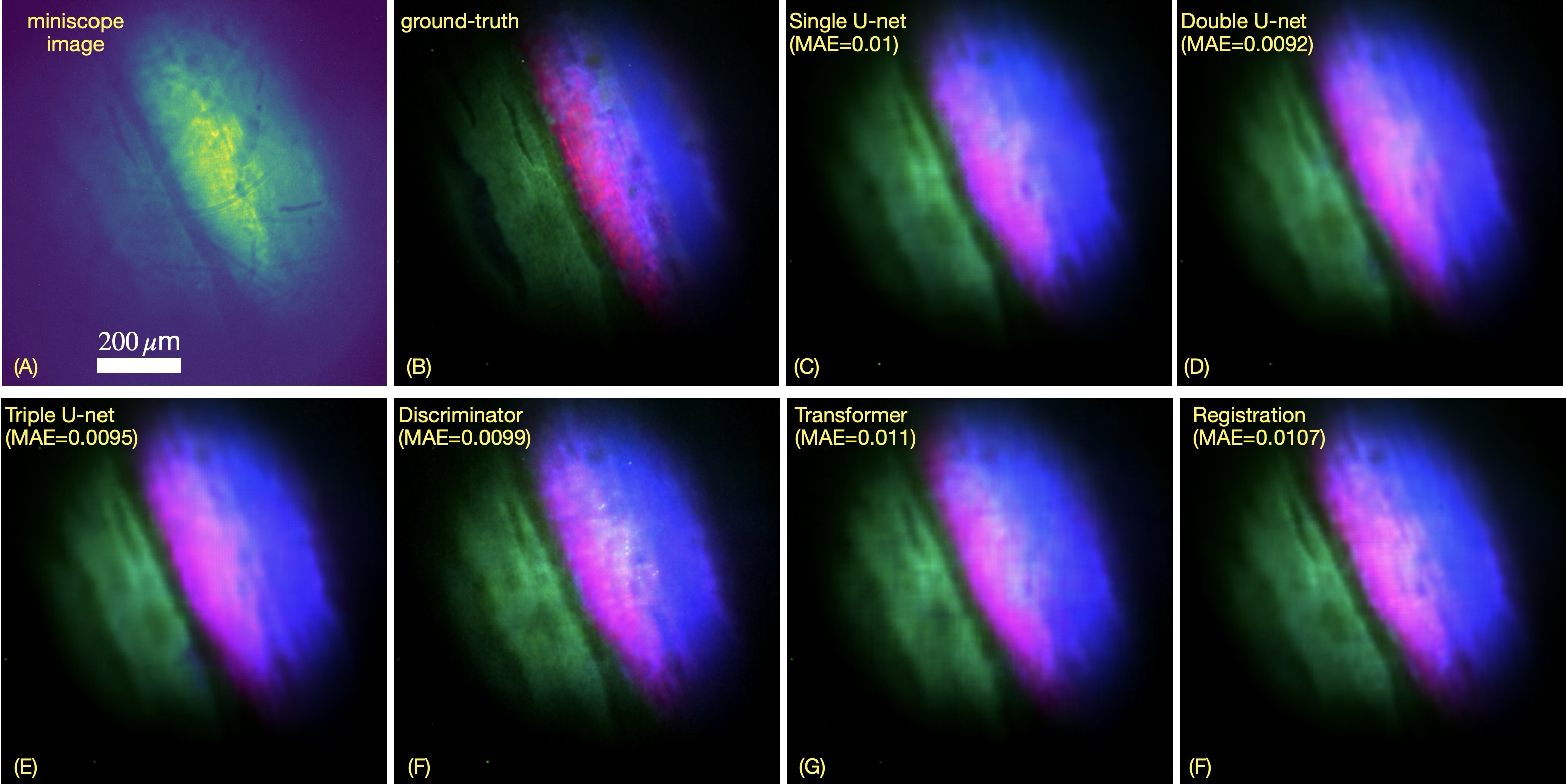}
\caption{Representative results for direct biomolecule mapping from one miniscope image, without intermediate hyperspectral reconstruction. (A) Recorded HAMscope image. (B) Ground-truth biomolecule map derived from the corresponding ground-truth hyperspectral images. (C–F) Predicted biomolecule maps using deep learning models trained on paired HAMscope and ground-truth biomolecule-mapping data. Mean absolute error (MAE) is indicated for each model.}\label{direct_mapping}
\end{figure}

The HAMscope encodes spectral information directly into each snapshot, allowing biomolecular distributions to be inferred from single-frame measurements without requiring full hyperspectral reconstruction. To generate reference maps for supervised training and evaluation, we applied the validated spectral profiles from Fig. \ref{spectral_unmixing}A to all 30-channel hyperspectral image stacks, producing spatial maps of lignin, chlorophyll, and residual autofluorescence components. An example ground-truth decomposition is shown in Fig. \ref{direct_mapping}B, corresponding to the recorded image in Fig. \ref{direct_mapping}A. These maps were used as targets for training models to directly predict biomolecule distributions from encoded snapshots. 

Representative outputs from multiple network architectures are shown in Figs. \ref{direct_mapping}C–F, highlighting predictions from several network architectures trained to regress biomolecule distributions directly from raw HAMscope images. All tested biomolecule-mapping models achieved comparable performance, with mean absolute errors (MAE) $\sim$0.0064. Notably, the direct mapping approach yielded lower error than the two-step strategy of first predicting hyperspectral data followed by biomolecule unmixing (MAE $\sim$ 0.0327), suggesting that the network more effectively learns biomolecule-specific spectral features than full-spectrum reconstruction. Among probabilistic biomolecule-mapping models, a standard single U-Net achieved an MAE of 0.00101. U-nets with two to five passes achieved MAEs of 0.00924, 0.00953, 0.00923, and 0.00935, respectively (Table S1). All multi-pass U-nets exceeded the performance of a U-net with twice the number of filters per layer (from 64 to 128) and fourfold the parameter count (MAE of 0.00988). 

\subsection{Resolution}
\begin{figure}[htb!]
\centering
\includegraphics[width=\linewidth]{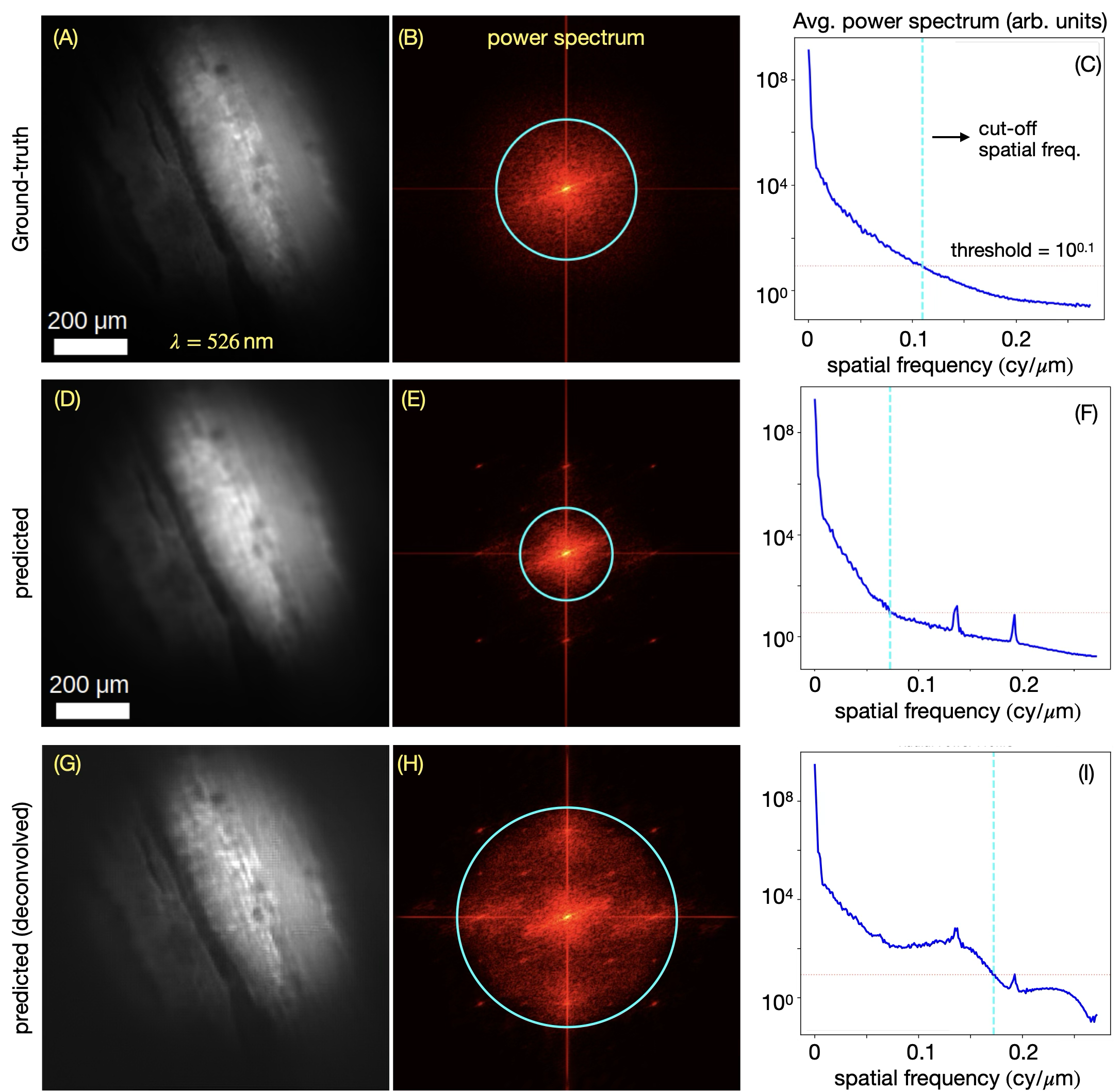}
\caption{Spatial resolution assessment of HAMscope reconstructions.
(A–C) Ground-truth image at 526 nm (A), corresponding 2D Fourier transform (B), and azimuthally averaged power spectrum (C). (D–F) Predicted image from the hyperspectral U-Net (D), its Fourier transform (E), and power spectrum (F). (G–I) Same spectral slice after space-variant Wiener deconvolution \cite{Ingold2025} (G), with corresponding Fourier transform (H) and power spectrum (I).
Spatial resolution was quantified by identifying the spatial frequency at which the log power spectrum dropped below a 0.1 threshold. Averaged over 100 test samples and 30 spectral channels, the network reconstructions achieved resolutions of 10.22 $\mu$m before and 6.74 $\mu$m after deconvolution, compared to 9.13 $\mu$m for the corresponding ground-truth image. Note that the ground-truth images were acquired using a different microscope configuration (objective and detector). Full resolution analysis is provided in Supplementary Fig. S10.}\label{resolution}
\end{figure}

To quantify the effective spatial resolution of the predicted spectral images, we performed a Fourier-based analysis that estimates the highest recoverable spatial frequency content. For each image, we computed the two-dimensional Fast Fourier Transform (FFT) and derived the corresponding power spectrum (Fig. \ref{resolution}). To reduce dimensionality and enable robust quantification, we converted the 2D spectrum into a one-dimensional radial profile by azimuthally averaging the power at increasing spatial frequency radii. The resolution cutoff was defined as the spatial frequency at which the logarithm of the radial power spectrum fell below a fixed threshold of 0.1, marking the transition from signal-dominated to noise-dominated frequencies. This cutoff frequency was then converted to a real-space resolution (in micrometers) by normalizing the pixel size and field of view. The resulting metric provides an objective and reproducible measure of image sharpness. Averaged across all spectral channels of 100 test images, the predicted spectral reconstructions achieved a wavelength-averaged spatial resolution of 10.22 $\mu$m. As expected, input images from the diffuser-based miniscope exhibited significantly degraded resolution (see Fig. S10).

To recover spatial resolution in our reconstructed hyperspectral images, we implemented a calibrated Wiener deconvolution approach. This method leverages a single calibration scan consisting of 100 randomly illuminated pixels to generate a dense map of point spread functions (PSFs) across all spectral channels \cite{Ingold2025, kohli2025ring}. The resulting calibration image was processed through the registration-trained single U-Net, producing a 30-channel hyperspectral PSF stack that captured network-induced spatial aberrations. Deconvolution using these PSFs yielded a marked improvement in resolution, with Fourier domain analysis indicating a mean lateral resolution of 6.74 $\mu$m across channels (Fig. \ref{resolution}C). Additional validation using a USAF resolution target is provided in Supplementary Fig. S11. For clarity, deconvolution is not applied to any other images in this work.

\subsection{Autofluorescence video imaging of wound response and embolism}
\begin{figure}[htb!]
    \centering
    \includegraphics[width=0.75\linewidth]{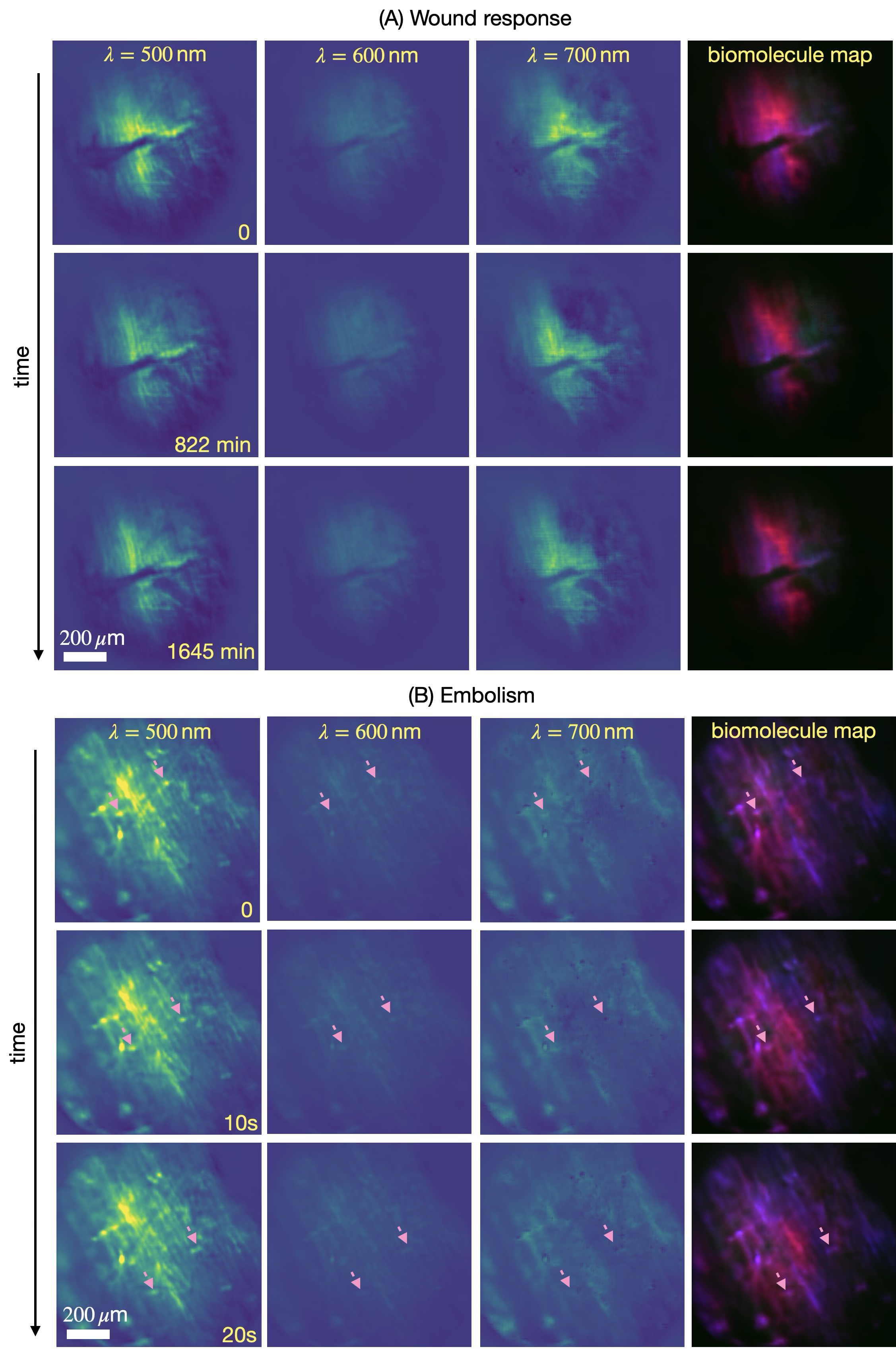}
    \caption{HAMscope video imaging of (A) time-lapse of wound response and (B) real-time embolisms in a transgenic poplar plant sample (Supplementary videos 1-4). For clarity, three representative spectral channels and three time points are shown. The corresponding biomolecule maps derived from the spectral data are displayed in the rightmost panels, highlighting dynamic molecular changes during tissue recovery. The arrows indicate two distinct emboli as they travel down the vessel elements.}
    \label{wound_embolism}
\end{figure}

To image internal tissues beneath the opaque outer surface of a living plant, we introduced a minimally invasive incision to enable direct optical access. The required opening for the miniscope was small (3 $\times$ 6 mm), allowing rapid tissue recovery post-incision. Using this preparation, we recorded time-lapse sequences of the wound-healing response in poplar stem over a period of 43 hours. The HAMscope is capable of real-time imaging at 10 frame/s, while the ground-truth microscope required 141 s for each hyperspectral frame. The biomolecule mapping framework described earlier (see Fig. \ref{direct_mapping}) was applied to these datasets to generate temporally resolved maps of autofluorescent biomarkers. Representative results from three spectral channels across three time points are shown in Fig. \ref{wound_embolism}A, with corresponding biomolecule maps in the rightmost panels (corresponding ground-truth images are in Fig. S13). The following were identified: lignin (blue), chlorophyll (green), and other autofluorescence (red). We observed that the red autofluorescence of chlorophyll decreases over time (Fig. S12). Complete time-lapse videos are provided in Supplementary Videos 1-4. 

Woody tissues are composed of xylem vessels, which—analogous to blood vessels—transport water and nutrients from roots to leaves. These conduits operate under negative hydraulic pressure, enabling the upward pull of water. However, this tension also renders xylem vulnerable to emboli, wherein air enters and displaces water within the vessel lumen. Embolism formation, often triggered by drought stress or mechanical injury, compromises vascular function and is rarely reversed naturally. Using the HAMscope, we captured real-time imaging of multiple embolism events induced by a mechanical incision through a transgenic poplar sample (Fig. \ref{wound_embolism}B and Supplementary Videos 5, 6). We note that observing such emboli dynamically is generally rare and challenging. This experiment reveals the potential of our microscope to investigate the spatio-spectral-temporal dynamics of embolism onset and propagation.

\subsection{Transgenic Poplar Imaging}
Stem segments from hybrid aspen ({\it Populus tremula} $\times$ {\it P. alba}, clone INRA 717-1B4) were excised under sterile conditions. Transgenic plants producing 2-Pyrone-4,5-dicarboxylic acid (denoted as PDC) and wild-type (WT) plants were compared. PDC is a biodegradable monomer for high-performance plastics. Engineering its biosynthetic pathway into plants enables PDC accumulation of up to ~3\% dry weight, while simultaneously reducing lignin content and enhancing biomass digestibility without affecting growth \cite{lin2021planta}. Clonal propagation was performed on half-strength Murashige and Skoog (½ MS) medium supplemented with 0.5 $\mu$M indole-3-butyric acid (IBA) for 6 weeks. Regenerated plantlets were transferred to soil and grown under controlled greenhouse conditions (16 h light / 8 h dark photoperiod at 25$^{\circ}$C) for an additional two months prior to stem harvest. Mature stems were then harvested and prepared for imaging. 

To assess biochemical differences across genotypes, we performed hyperspectral autofluorescence imaging at 65 discrete sites spanning tangential incisions on WT and PDC transgenic stems. For this dataset, a filter-wheel-based hyperspectral system was employed in place of the LVBF configuration. Emission spectra were acquired using a six-position motorized filter wheel (Thorlabs FW102C) with 40 nm bandpass filters centered at 400, 450, 500, 550, 600, and 650 nm (Thorlabs FBH series).

\begin{figure}[htb!]
    \centering
    \includegraphics[width=0.85\linewidth]{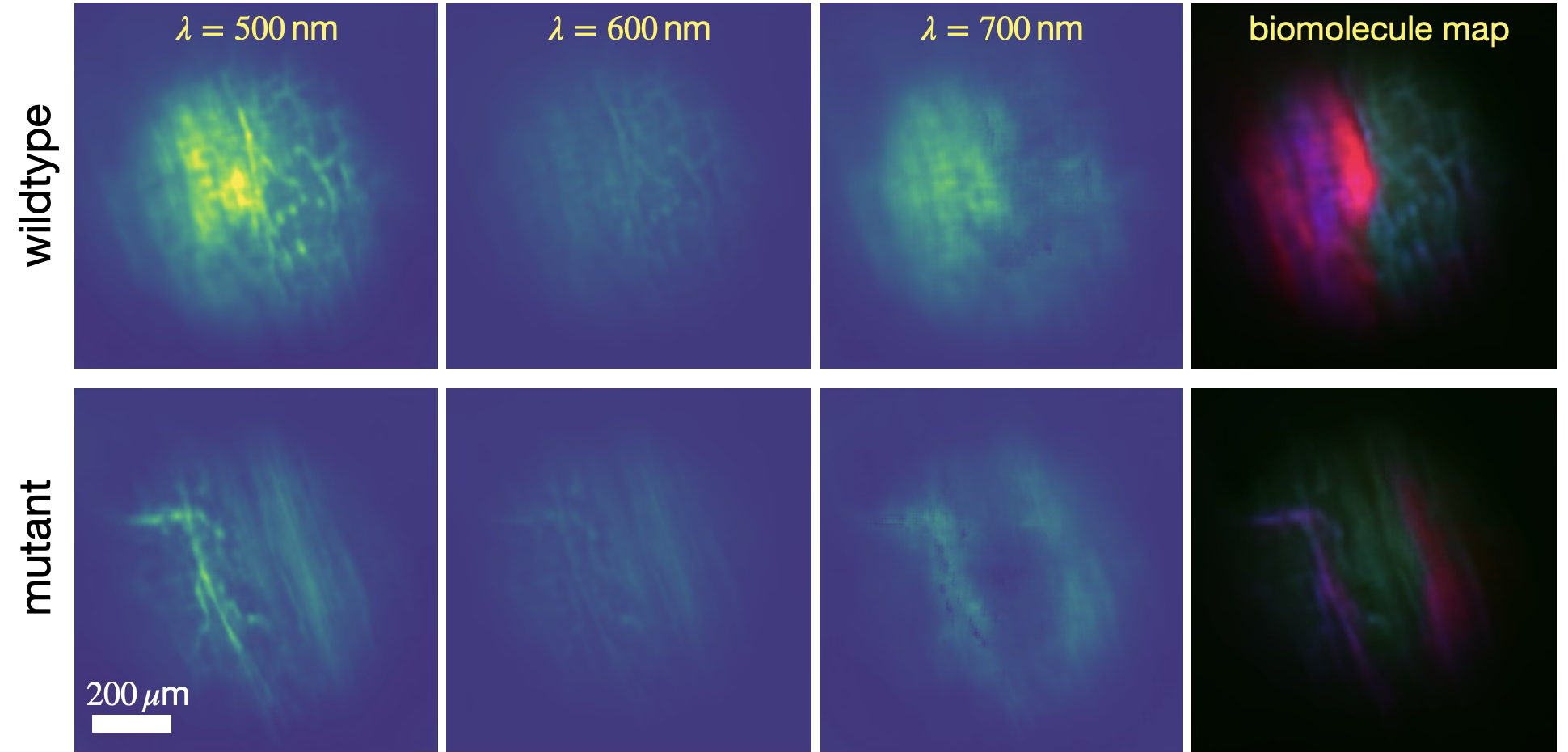}
    \caption{HAMscope imaging of wild-type and lignin-deficient PDC transgenic poplar stems. Hyperspectral imaging was performed at 65 sites across tangential incisions on wild-type (WT) and PDC transgenic Poplar branches using the HAMcope. Top: A representative autofluorescence image of xylem tissue in a WT branch acquired at 500 nm (40 nm bandwidth) shows a strong lignin-associated signal. Bottom: Corresponding image from a PDC transgenic sample, exhibiting reduced fluorescence intensity consistent with diminished lignin content.}
    \label{transgenic}
\end{figure}

Spectral analysis revealed a consistent reduction in lignin-associated autofluorescence in PDC transgenic stems relative to wild-type controls. This difference was most pronounced in the 500–550 nm range (Fig. \ref{transgenic}, corresponding ground-truth in Fig. S14), where lignin typically exhibits strong autofluorescence. Across 65 sites, transgenic samples displayed significantly lower fluorescence intensities in xylem tissues, consistent with their genetically reduced lignin content. Representative 500 nm emission images further highlight the diminished xylem signal in PDC lines compared to WT (see Fig. S14 for ground-truth images). The wild type and transgenic poplar with the hyperspectral registration network achieved an average hyperspectral MAE (biomolecular-mapping MAE) of 0.00552 (0.0113) and 0.00564 (0.0158) over 65 images, respectively. No ground truth images were available for this experiment. For reference, the test dataset of poplar incisions has hyperspectral MAE and biomolecular-mapping MAE 0.00457 and 0.00934, respectively. This marginal increase in MAE indicates high accuracy. The decreased scale (see Table S2) in the transgenic poplar can be attributed to decreased average fluorescence intensity from the PDC modification.  Averaging 65 wild-type and 65 PDC poplars, we observed lower autofluorescence in the 450-550 nm lignin range in the transgenic low-lignin poplars compared to the controls. This finding held true for ground truth and reconstructed images, even after the results were normalized (Fig. S15). This experiment used a six-channel filter wheel to acquire ground-truth emission data (in contrast to the LVBF-based setup used elsewhere in this study), but 30-channel data was generated by the HAMscope.

\subsection{Suberin Identification}
\begin{figure}[htb!]
    \centering
    \includegraphics[width=0.7\linewidth]{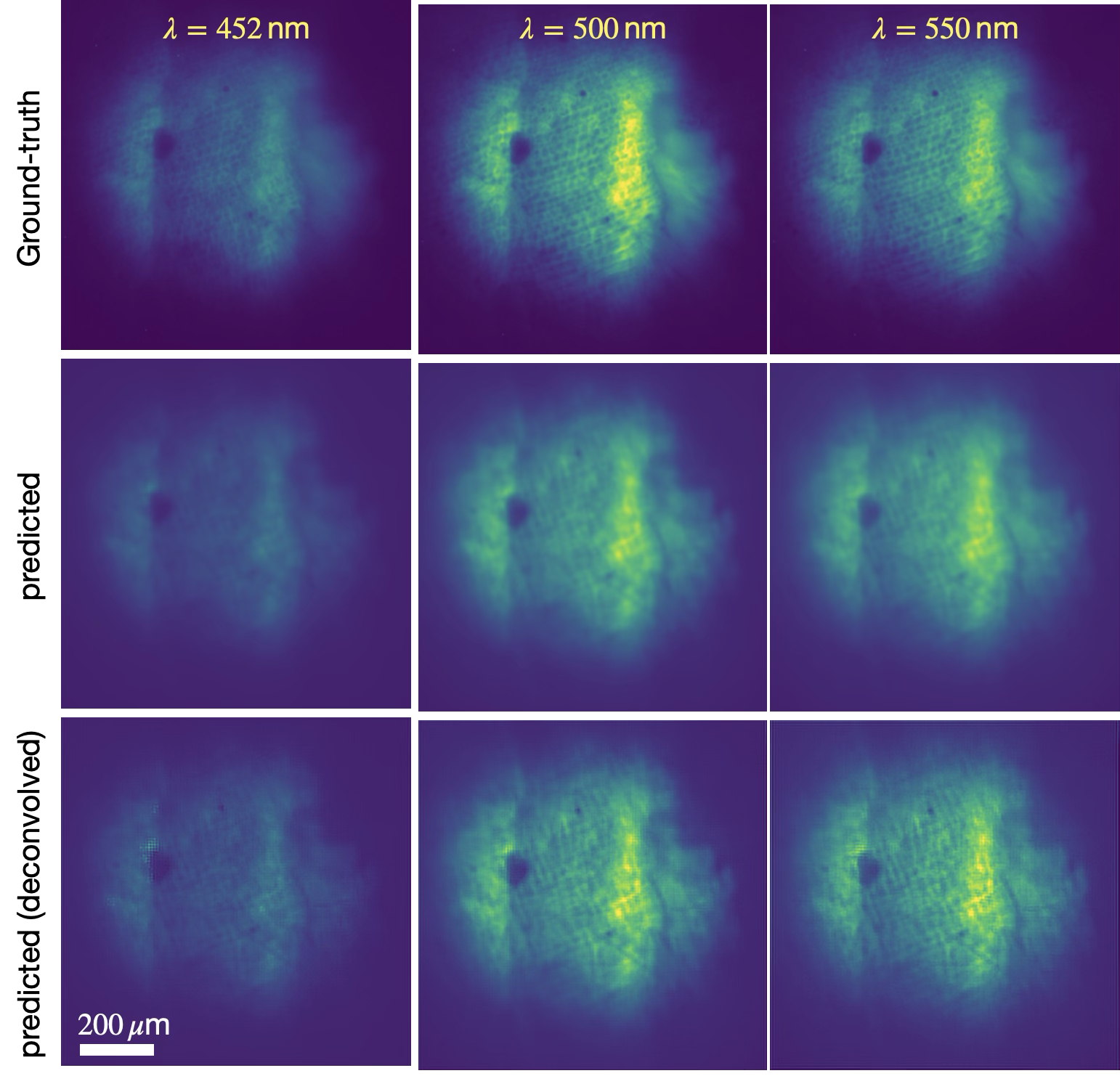}
    \caption{HAMscope imaging of suberin in cork. Exemplary predicted spectral channels along with the corresponding ground-truth and deconvolved images are shown. Registration and single U-net model was used for these images.}
    \label{suberin}
\end{figure}

Suberin is another important plant cell wall polymer that functions to protect plants from water loss\cite{de2021root,gracca2007suberin}. To test the HAMscope’s ability to detect suberin, we used the bark of {\it Quercus suber} (cork oak), which is a commercially valuable source of cork, composed primarily of 39\% suberin and 21\% lignin \cite{Pereira1988}. These distinct biochemical features—particularly elevated suberin and reduced chlorophyll levels—place the sample distribution well outside the poplar-based training set, providing a challenging test case for the HAMscope. Representative HAMscope-generated images are shown in Fig. \ref{suberin}, with corresponding ground-truth and scale hyperspectral images provided in Fig. S16. A custom biomolecular mapping approach was used to separate the autofluorescence contributions of lignin, suberin, and chlorophyll only using the ground-truth images (Fig. S17). Imaging of suberin using the hyperspectral registration network achieved a mean absolute error (MAE) of 0.0718 and an average scale value of 0.00721 across 100 images. Unlike poplar, which exhibits a narrow red autofluorescence band (650–700 nm) due to chlorophyll, cork oak bark shows minimal chlorophyll fluorescence. This spectral mismatch led to a systematic overestimation of the 700 nm band, which is effectively flagged by high uncertainty values in the corresponding scale image.

\subsection{Comparing neural network models}
\begin{figure}[htb!]
\centering
\includegraphics[width=\textwidth]{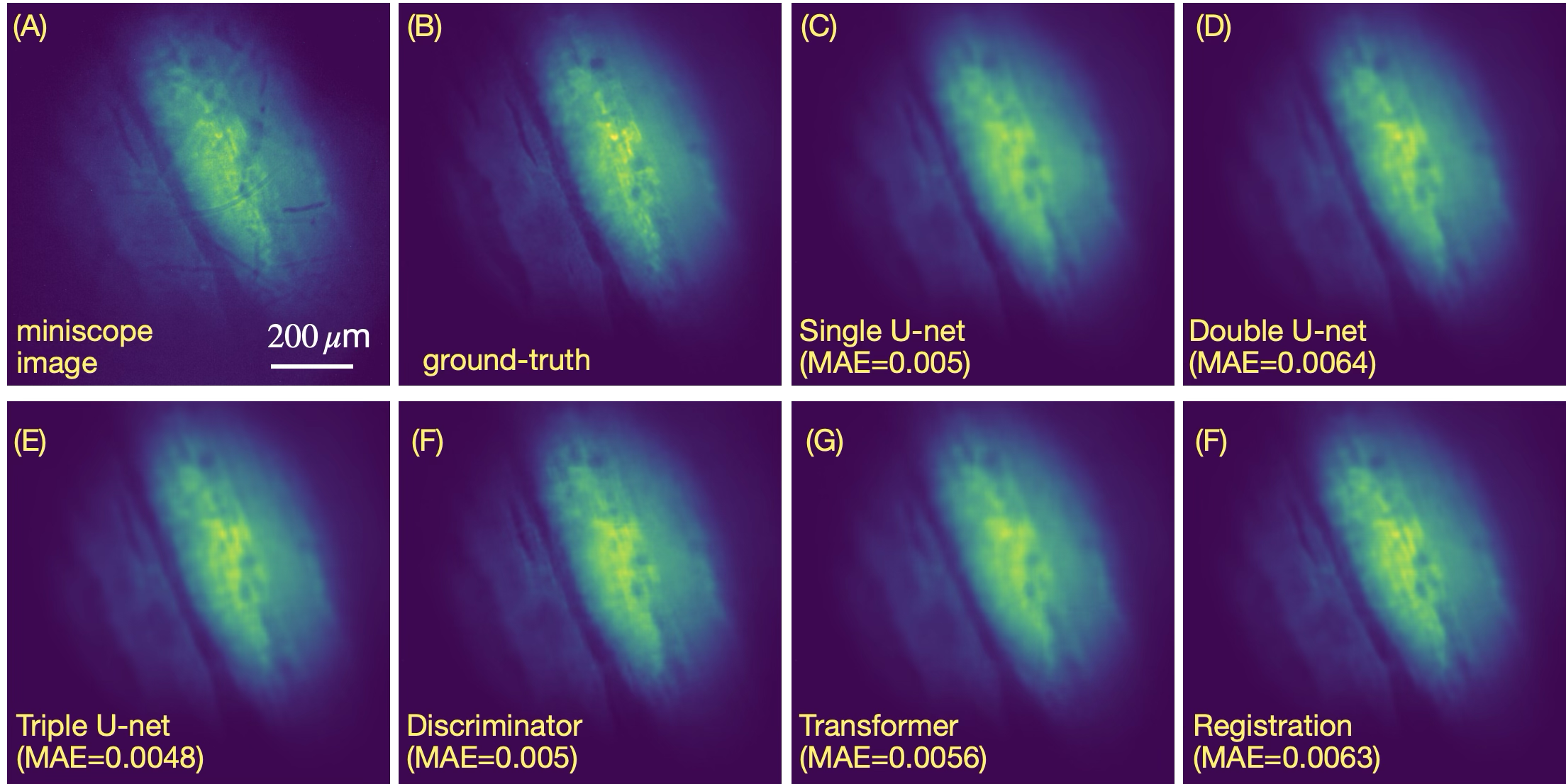}
\caption{Comparison of Various Models Used. Mean Average Error (MAE) is displayed for the 100 randomly selected test images. The 200 $\mu$m scalebar represents all images.}\label{Model_comparison}
\end{figure}

Reconstruction accuracy was assessed using mean absolute error (MAE) on a held-out test set of 100 images. Among the probabilistic hyperspectral models, a standard single-pass U-Net achieved an average MAE of 0.00496, based on an ensemble of seven identical models (standard deviation: 0.000058). A dual-pass (double U-Net) architecture yielded a slightly higher MAE of 0.00640, while further increasing depth to a triple-pass U-Net improved performance, achieving the lowest MAE of 0.0048 (Fig. \ref{Model_comparison}). Architectural variations provided incremental gains, with the triple U-Net consistently outperforming other configurations. Notably, a model incorporating image registration performed particularly well, though its MAE (0.00607) likely underestimates perceptual quality due to uncorrected residual misalignments in the test set. Given its qualitative advantages, we selected the registration-augmented model for downstream analyses shown in Figs. \ref{fig:miniscope}, \ref{spectral_unmixing}, \ref{resolution}, \ref{wound_embolism}, \ref{transgenic}, and \ref{suberin}. The single U-net was used in fig. \ref{HSI_core_results}.

We observed minimal variability across replicate training runs for the base probabilistic U-Net (standard deviation of MAE = 0.0001), whereas models incorporating discriminators, transformers, or registration exhibited greater variance. Among these, the discriminator-based model consistently underperformed under the current normalization scheme (scaling by sensor bit depth). However, in earlier experiments using min–max normalization, the discriminator yielded the best results, highlighting sensitivity to preprocessing. Transformer-augmented models showed marginally worse performance than the base U-Net and incurred a substantial increase in training time. The double U-Net variant performed comparably to the single U-Net, while the triple U-Net outperformed all others in terms of MAE. Full quantitative metrics, including MAE, MSE, and SSIM for all models, are reported in Supplementary Table 1. Ablation results for models trained without the probabilistic Laplacian NLL loss are shown in Supplementary Fig. S3.

\section{Conclusions}
In this study, we introduce the HAMscope, a compact, snapshot hyperspectral autofluorescence microscope that transforms a widely used miniscope platform into a high-throughput biochemical imaging tool suitable for in situ plant studies. By integrating a thin polymer diffuser at the image plane and employing a deep learning pipeline, we achieve real-time hyperspectral imaging across 30 spectral channels without compromising the spatial resolution, frame rate, or physical footprint of the original device. The resulting system directly maps endogenous biomolecules—such as lignin, chlorophyll, and suberin—at cellular resolution, even within living plant tissues in field-like conditions. Although the initial demonstrations focused on plant systems, the core imaging and computational approach is fully general and readily extendable to other biological contexts, including neuroscience, metabolic imaging, and intraoperative diagnostics.

Beyond image reconstruction, we developed a probabilistic deep learning framework that directly predicts spectral profiles and compound-specific abundance maps from raw sensor images. By estimating pixel-wise uncertainties, the model improves accuracy and interpretability relative to conventional spectral unmixing approaches. Notably, we show that direct biomolecular mapping from encoded single-shot measurements can exceed the fidelity of traditional two-step reconstruction pipelines, offering a new paradigm for low-latency imaging. The framework supports modular components—including a multi-pass U-Net generator, probabilistic loss functions, a transformer-based attention module, a discriminator for adversarial training, and an image registration block. Multi-pass models stood out with their accuracy and efficiency while the registration model preserved detail through automatic alignment. While no individual module consistently outperformed the baseline across all datasets, each contributed superior performance at specific stages of development. This unified, pix2pix-inspired architecture provides a flexible platform for researchers to optimize image-to-image translation tailored to their own datasets and applications. Importantly, the software architecture is domain-agnostic and can be trained on paired or synthetic data from any samples, enabling widespread adoption beyond plant sciences.

While HAMscope shows strong performance across multiple applications—from transgenic phenotyping to embolism dynamics—it inherits some limitations from its underlying hardware. These include constrained optical throughput and modest spatial resolution ($\sim10\thinspace\mu$m), which can be improved in the future by using custom-designed miniaturized objective lenses. Additionally, current training pipelines rely on carefully synchronized dual-path setups for supervision. Future work will investigate self-supervised and transfer learning frameworks to relax the need for paired ground-truth and broaden applicability to new tissues and species.

Looking ahead, we envision HAMscope as a foundation for distributed biochemical imaging across ecosystems, greenhouses, and crop fields. Extensions to additional excitation wavelengths, incorporation of microendoscopic probes, and integration with edge AI hardware could further expand its reach. More broadly, HAMscope offers a general-purpose solution for low-cost, real-time, spectral imaging in miniaturized systems, with potential applications in neuroscience (e.g., brain metabolism and vasculature), biomedical diagnostics, wearable health monitoring, and point-of-care pathology. The principles demonstrated here—snapshot spectral encoding, direct molecular mapping, and uncertainty-aware inference—may find applications in neural imaging, pathology, and portable diagnostics where label-free biochemical contrast is essential.

\section*{Methods}
\subsection*{Dataset Acquisition Software}
Automated data acquisition was implemented in LabVIEW (version 16.0f5, 32-bit), enabling high-throughput capture of thousands of images per dataset. Each acquisition cycle consisted of six sequential ground-truth images, followed by one image from the miniscope. LabVIEW controlled the motorized variable bandpass filter and dichroic mirror, triggering filter changes between each ground-truth frame and coordinating mirror movements before and after each miniscope capture. After completing each acquisition cycle, the sample stage (Thorlabs, motorized XYZ) was advanced to the next location in a predefined 3D scan path. The stage moved in 100 $\mu$m increments along a curved surface, with the axial (z) height programmed to follow the contours of the poplar branch. Surface geometry was defined by manually placing support points at a density of approximately one per mm², from which the full 3D surface was interpolated. The stage perimeter was similarly defined by fitting a smooth boundary around the support points. Ground-truth images were acquired directly via the Hamamatsu Video Capture Library (version 4418), which is natively supported in LabVIEW. In contrast, the miniscope software lacked native integration; image acquisition was triggered via a LabVIEW-controlled auto-clicker interface to the open-source capture tool\cite{AharoniLab2024MiniscopeDAQ}.

\subsection*{Sample preparation}
For training and testing of the primary dataset, branch cuttings from one-year-old \textit{Populus Nigra × Deltoides} trees were potted in Sunshine Mix \#4 (Sun Gro Horticulture) and maintained in a greenhouse for two years under a 14 h light / 10 h dark photoperiod. For an image of the poplars in the greenhouse, see Fig. S3 E. Imaging samples were prepared from branches at varying developmental stages, specifically between the 5th and 10th internodes. A narrow strip of bark ($\sim$3 cm in length, $\sim$200 $\mu$m in depth) was carefully removed using a miniature wood plane (Veritas Miniature Bench Plane, Lee Valley Tools) to expose the underlying xylem, as shown in Fig. \ref{spectral_unmixing}B for a training dataset incision or Fig. S3C for a smaller test incision. Branches were stabilized during imaging by mounting them on a tip-tilt stage (KM100B, Thorlabs) using a custom-designed 3D-printed clamp (see code availability). 

\begin{backmatter}
\bmsection{Funding}
Funding from US Dept. of Energy grant \#55801063 is gratefully acknowledged. CJL and ND were supported by the Joint BioEnergy Institute, one of the Bioenergy Research Centers of the US DOE, Office of Science, Office of Biological and Environmental Research, through contract DE-AC02-05CH11231 between Lawrence Berkeley National Laboratory and the US DOE.

\bmsection{Acknowledgment}
We are grateful for discussions with Andrew Groover, Apratim Majumder, Ryan Manwill, and Aadhi Umamageswaran.

\bmsection{Disclosures}
\noindent The authors declare no competing interests.

\bmsection{Data Availability Statement}
Data underlying the results presented in this paper are not publicly available at this time but may be obtained from the authors upon reasonable request.

\bmsection{Supplemental document}
See Supplement 1 for supporting content.
\end{backmatter}

%%%%%%%%%%%%%%%%%%%%%%% References %%%%%%%%%%%%%%%%%%%%%%%%%

%%%%%%%%%% If using BibTeX:
\bibliography{sample}

\end{document}